\documentclass[aps,floatfix,prd,superscriptaddress,reprint,showpacs,amsmath,amssymb,longbibliography,nofootinbib]{revtex4-2}

\usepackage{amsmath,amssymb}
\usepackage[dvipsnames]{xcolor}
\usepackage[pdftex,
  pdftitle={The eta->g*g* transition form factor and the hadronic light-by-light eta-pole contribution to the muon g-2 from lattice QCD},
  pdfauthor={C. Alexandrou, S. Bacchio, S. Burri, J. Finkenrath,
    A. Gasbarro, K. Hadjiyiannakou, K. Jansen, G. Kanwar, B. Kostryewa,
    K. Ottnad, M. Petschlies, F. Pittler, C. Urbach, U. Wenger},
  bookmarks,
  colorlinks,
  linkcolor=blue,
  citecolor=blue,
  filecolor=blue,
  urlcolor=blue,
  menucolor=black,
  plainpages=false,
  pdfpagelabels,
  hypertexnames=false]{hyperref}
\usepackage[capitalise]{cleveref}
\usepackage{braket}
\usepackage[utf8]{inputenc}
\usepackage{graphicx}
\usepackage{mathtools}

\usepackage{tikz,tikz-feynman}
\tikzfeynmanset{compat=1.1.0, warn luatex=false}
\usetikzlibrary{arrows, arrows.meta}

\newcommand{\eps}[0]{\epsilon}

\newcommand{\fig}[0]{Fig.}

\newcommand{\vvTFF}[1]{\mathcal{F}_{#1 \rightarrow \gamma^* \gamma^*}}
\newcommand{\vrTFF}[1]{\mathcal{F}_{#1 \rightarrow \gamma^* \gamma}}

\newcommand{\rrTFF}[1]{\mathcal{F}_{#1 \rightarrow \gamma \gamma}}
\newcommand{\amu}[1]{a_\mu^{#1\text{-pole}}}

\newcommand{\ampl}[0]{\tilde{A}}

\DeclareMathOperator{\diag}{diag}

\begin{document}

\graphicspath{{./figs/}{./supp_figs/}}
\makeatletter
\def\input@path{{./figs/}{./supp_figs/}}
\makeatother

\title{The \texorpdfstring{$\eta \rightarrow \gamma^* \gamma^*$}{eta -> g* g*} transition form factor and the hadronic light-by-light \texorpdfstring{$\eta$-pole}{eta-pole} contribution to the muon \texorpdfstring{$g-2$}{g-2} from lattice QCD}

\author{Constantia Alexandrou}
\affiliation{Department of Physics, University of Cyprus, Nicosia, Cyprus}
\affiliation{Computation-based Science and Technology Research Center, The Cyprus Institute, Nicosia, Cyprus}

\author{Simone Bacchio}
\affiliation{Computation-based Science and Technology Research Center, The Cyprus Institute, Nicosia, Cyprus}

\author{Sebastian Burri}
\affiliation{Albert Einstein Center, Institute for Theoretical Physics, University of Bern, Switzerland}

\author{Jacob Finkenrath}
\affiliation{Computation-based Science and Technology Research Center, The Cyprus Institute, Nicosia, Cyprus}

\author{Andrew Gasbarro}
\affiliation{Albert Einstein Center, Institute for Theoretical Physics, University of Bern, Switzerland}

\author{Kyriakos Hadjiyiannakou}
\affiliation{Department of Physics, University of Cyprus, Nicosia, Cyprus}
\affiliation{Computation-based Science and Technology Research Center, The Cyprus Institute, Nicosia, Cyprus}

\author{Karl Jansen}
\affiliation{NIC, DESY Zeuthen, Germany}

\author{Gurtej Kanwar}
\affiliation{Albert Einstein Center, Institute for Theoretical Physics, University of Bern, Switzerland}

\author{Bartosz Kostrzewa}
\affiliation{High Performance Computing and Analytics Lab, Rheinische Friedrich-Wilhelms-Universit\"at Bonn,
Germany}

\author{Konstantin Ottnad}
\affiliation{PRISMA$^+$ Cluster of Excellence and Institut f\"ur
Kernphysik, Johannes Gutenberg-Universit\"at Mainz, Germany}

\author{Marcus Petschlies}
\affiliation{Helmholtz-Institut f\"ur Strahlen- und Kernphysik, University of Bonn, Germany}
\affiliation{Bethe Center for Theoretical Physics, University of Bonn, Germany}

\author{Ferenc Pittler}
\affiliation{Computation-based Science and Technology Research Center, The Cyprus Institute, Nicosia, Cyprus}

\author{Carsten Urbach}
\affiliation{Helmholtz-Institut f\"ur Strahlen- und Kernphysik, University of Bonn, Germany}
\affiliation{Bethe Center for Theoretical Physics, University of Bonn, Germany}

\author{Urs Wenger}
\affiliation{Albert Einstein Center, Institute for Theoretical Physics, University of Bern, Switzerland}

\collaboration{Extended Twisted Mass Collaboration}

\newcommand{\amuresult}
{13.75(5.24)_\text{stat}(1.53)_\text{syst}}
\newcommand{\amuresultshortNoTot}
{13.8(5.2)_\text{stat}(1.5)_\text{syst}}
\newcommand{\amuresultshort}
{\amuresultshortNoTot[5.5]_\text{tot}}
\newcommand{\GammaresultNoTot}
{338(87)_\text{stat}(17)_\text{syst}}
\newcommand{\Gammaresult}
{\GammaresultNoTot[88]_\text{tot}}
\newcommand{\bPresultNoTot}
{1.34(28)_\text{stat}(14)_\text{syst}}
\newcommand{\bPresult}
{\bPresultNoTot[31]_\text{tot}}
\newcommand{\bPresultExpt}
{1.92(4)}
\newcommand{\bPresultDisp}
{1.95(9)}

\date{\today}

\begin{abstract}
 We calculate the double-virtual $\eta \rightarrow \gamma^* \gamma^*$ transition form factor $\vvTFF{\eta}(q_1^2,q_2^2)$ from first principles using a lattice QCD simulation with $N_f=2+1+1$ quark flavors at the physical pion mass and at one lattice spacing and volume. The kinematic range covered by our calculation is complementary to the one accessible from experiment and is relevant for the $\eta$-pole contribution to the hadronic light-by-light scattering in the anomalous magnetic moment $a_\mu = (g-2)/2$ of the muon. From the form factor calculation we extract the partial decay width $\Gamma(\eta \rightarrow \gamma \gamma) = \GammaresultNoTot$ eV and the slope parameter $b_\eta=\bPresultNoTot$ GeV${}^{-2}$.
 For the $\eta$-pole contribution to $a_\mu$ we obtain $\amu{\eta} = \amuresultshortNoTot \cdot 10^{-11}$.
\end{abstract}

\maketitle

\section{Introduction}
Radiative transitions and decays of the neutral pseudoscalar mesons $P=\pi^0$, $\eta$ and $\eta'$ arise through the axial anomaly and are therefore a crucial probe of the nonperturbative low-energy properties of QCD. 
The simplest transition to two (virtual) photons, $P \rightarrow \gamma^*\gamma^*$, is specified through the transition form factor (TFF) $\vvTFF{P}(q_1^2,q_2^2)$ defined by the matrix element
\begin{multline} \label{eq:tff-defn}
  i\int d^4x \, e^{iq_1 x} \langle 0 | T\{j_\mu(x) j_\nu(0)\} | P(q_1+q_2) \rangle \\
  = \epsilon_{\mu\nu\rho\sigma}q_1^\rho q_2^\sigma \vvTFF{P}(q_1^2,q_2^2),
\end{multline}
where $j_\mu, j_\nu$ are the electromagnetic currents and $q_1,q_2$ are the photon momenta. The TFFs determine the partial decay widths to leading order in the fine-structure constant $\alpha_\text{em}$  through
\begin{equation}
  \Gamma(P\rightarrow \gamma\gamma) = \frac{\pi \alpha_\text{em}^2 m_P^3}{4} \, |\rrTFF{P}(0,0)|^2,
\end{equation}
where $m_P$ is the pseudoscalar meson mass.
$\Gamma(\eta\rightarrow \gamma\gamma)$ is of particular interest, 
since it can be used to extract the $\eta-\eta'$ mixing angles 
and provides a normalization for many other $\eta$ partial widths~\cite{Workman:2022ynf}. At the same time, there is a long-standing tension between its 
different experimental determinations through $e^+e^-$ collisions on the one hand and Primakoff production on the other~\cite{Gan:2020aco,JADE:1985biu,CrystalBall:1988xvy,Roe:1989qy,Baru:1990pc,KLOE-2:2012lws,Browman:1974sj}.
The TFFs also provide input for determining the electromagnetic interaction radius of the pseudoscalar mesons through the slope parameter
\begin{equation}
  b_P = \left. \frac{1}{\rrTFF{P}(0,0)}\frac{\text{d}\vrTFF{P}(q^2,0)}{\text{d}q^2}\right|_{q^2=0} \, .
\end{equation}

\begin{figure}
    \centering
    \includegraphics{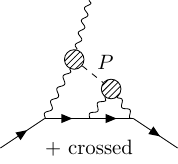}
    \hspace{0.2cm}
    \includegraphics{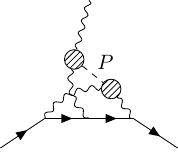}
    \caption{The pseudoscalar pole diagrams contributing to the
      leading order HLbL scattering in the muon anomalous magnetic
      moment. Each striped blob indicates the insertion of a pseudoscalar meson transition form factor $\vvTFF{P}$, where $P \in \{\pi^0, \eta, \eta'\}$.}
    \label{fig:ps-pole}
\end{figure}

Moreover, the TFFs play a critical role for the leading-order hadronic light-by-light (HLbL) scattering in the anomalous magnetic moment
$a_\mu = (g-2)/2$ of the muon.
Recent results from the Fermilab E989 and Brookhaven E821 experiments~\cite{Muong-2:2021ojo,Muong-2:2006rrc} indicate a $4.2\sigma$ tension with the consensus
on the Standard Model (SM) prediction in Refs.~\cite{Aoyama:2020ynm,Aoyama:2012wk,Aoyama:2019ryr,Czarnecki:2002nt,Gnendiger:2013pva,Davier:2017zfy,Keshavarzi:2018mgv,Colangelo:2018mtw,Hoferichter:2019mqg,Davier:2019can,Keshavarzi:2019abf,Kurz:2014wya,Melnikov:2003xd,Masjuan:2017tvw,Colangelo:2017fiz,Hoferichter:2018kwz,Gerardin:2019vio,Bijnens:2019ghy,Colangelo:2019uex,Blum:2019ugy,Colangelo:2014qya}. The uncertainty of the latter is dominated by the Hadronic Vacuum Polarization 
and the HLbL scattering. In particular, matching the planned improvement on the experimental uncertainty by a factor of four in the SM evaluation, an improved control of the uncertainty of the HLbL contribution is mandatory, cf.~Ref.~\cite{Aoyama:2020ynm}.
The HLbL contribution can be estimated, among other approaches~\cite{Kinoshita:1984it,deRafael:1993za,Bijnens:1995cc,Bijnens:1995xf,Bijnens:2001cq,Hayakawa:1995ps,Hayakawa:1996ki,Hayakawa:1997rq,Knecht:2001qf,Melnikov:2003xd}, by a systematic decomposition
into contributions from various intermediate states~\cite{Colangelo:2014dfa,Colangelo:2014pva,Colangelo:2015ama,Pauk:2014rfa}.
Lattice QCD can provide ab-initio data for the required form factors and hadron scattering amplitudes within this approach.
This is thus complementary to a lattice-QCD calculation of the full HLbL scattering amplitude~\cite{Blum:2014oka,Blum:2015gfa,Blum:2016lnc,Blum:2017cer,Chao:2021tvp,Chao:2022xzg}.

The pseudoscalar pole diagrams, depicted in \fig~\ref{fig:ps-pole}, make the dominant contribution to the HLbL scattering amplitude, with
$\vvTFF{P}$ as the key nonperturbative input. Of these diagrams, the
$\pi^0$-pole contribution has been estimated based on a dispersive
framework~\cite{Hoferichter:2018dmo,Hoferichter:2018kwz} and on lattice-QCD calculations of the pion TFF~\cite{Gerardin:2016cqj,Gerardin:2019vio,Burri:2021cxr} while
rational approximant fits to experimental TFF data have yielded
an estimate of all three contributions~\cite{Masjuan:2017tvw}.  A
preliminary calculation of the $\eta$- and $\eta'$-pole contributions using a coarse lattice
was reported in Ref.~\cite{Gerardin:2022jcw}. Experimental results from CELLO~\cite{CELLO:1990klc}, CLEO~\cite{CLEO:1997fho}, and BaBar~\cite{BaBar:2009rrj,BaBar:2011nrp} constrain the
spacelike single-virtual $\vrTFF{P}(-Q^2, 0)$ in the regime $Q^2 \gtrsim 1 \, \mathrm{GeV}^2$, but do not provide data for $0 \leq Q^2 \lesssim 1 \mathrm{GeV}^2$
or for general double-virtual kinematics.
In contrast, these kinematics are the most accessible in lattice
QCD and therefore provide highly relevant and important new information that is of interest for phenomenological models and various experimental efforts.

In this letter we present an ab-initio calculation of $\vvTFF{\eta}(q_1^2,q_2^2)$ and the corresponding $\eta$-pole HLbL contribution $\amu{\eta}$ using lattice QCD simulations at a
single lattice spacing and a single volume.  We employ $N_f=2+1+1$
flavors of twisted-mass quarks~\cite{Frezzotti:2000nk} tuned to the
physical pion mass, physical heavy-quark masses, and maximal
twist. The latter guarantees automatic
${\cal O}(a)$-improvement of observables~\cite{Frezzotti:2003ni,Frezzotti:2004wz},
which here includes $\vvTFF{P}$, $\Gamma(\eta \to \gamma\gamma)$, $b_\eta$, and $\amu{\eta}$. \\

\section{Methods}
We apply the method introduced in Refs.~\cite{Gerardin:2016cqj,Gerardin:2019vio} to the case of the $\eta$ TFF. Details of our analysis are specified below.

\subsection{Amplitude and kinematics}
In particular, the TFF is related to the Euclidean $\eta$-to-vacuum transition amplitude~\cite{Ji:2001wha} 
\begin{equation}
\ampl_{\mu\nu}(\tau) \equiv \int d^3\vec{x} e^{-i \vec{q}_1 \cdot \vec{x}} \braket{0 | T\{ j_\mu(\tau,\vec{x}) j_\nu(0) \} | \eta(\vec{p})}
\label{eq:eucl-amplitude}
\end{equation}
by
\begin{equation} 
  \begin{aligned}
    \eps_{\mu\nu\alpha\beta} q_1^\alpha q_2^\beta \, \vvTFF{\eta}(q_1^2, q_2^2) = -i^{n_0} \hspace{-4pt}\int_{-\infty}^{\infty} \hspace{-6pt}d\tau e^{\omega_1 \tau} \hspace{-2pt}\ampl_{\mu\nu}(\tau),
\end{aligned}
\label{eq:tff-eucl-amplitude}
\end{equation}
where $n_0 = \delta_{\mu,0} + \delta_{\nu,0}$ counts the number of temporal indices.
\begin{figure}
    \centering
    \vspace{-0.1cm}
    \includegraphics{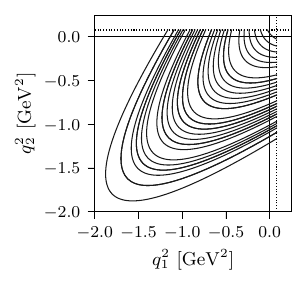}
    \vspace{-0.3cm}
    \caption{Orbits of photon virtualities $(q_1^2, q_2^2)$ accessed in this work. Dotted lines indicate two-pion thresholds at $4 m_\pi^2$.}
    \label{fig:eta-orbits}
\end{figure}
The kinematics are determined by the four-momentum $p \equiv (E_\eta, \vec{p}\,)$ of the on-shell $\eta$ state with energy $E_\eta = \sqrt{m_\eta^2 + \vec{p}\,^2}$, the four-momentum $q_1 = (\omega_1, \vec{q}_1)$ of the first current, and the momentum conservation constraint $q_2 = p - q_1$. In the lattice setup used here, it is most practical to fix $\vec{p}$ and evaluate the amplitude for a variety of $\vec{q}_1$ and $\omega_1$. The present calculation is restricted to the rest frame, $\vec{p} = (0,0,0)$, and momenta satisfying $|\vec{q}_1|^2 \leq 32 (2\pi / L)^2$ and $|q_1^{x}|, |q_1^{y}|, |q_1^{z}| \leq 4 (2 \pi / L)$. Each choice of finite-volume momentum $\vec{q}_1$ gives access to $\vvTFF{\eta}(q_1^2, q_2^2)$ on a particular kinematical orbit, as shown in \fig~\ref{fig:eta-orbits}. Notably, the $|\vec{q}_1|^2 = (2\pi / L)^2$ orbit lies outside the spacelike quadrant, but still falls below the nonanalytic thresholds at $4 m_\pi^2$, allowing it to be accessed on the lattice; its proximity to $(0,0)$ makes it particularly helpful in constraining $\Gamma(\eta \to \gamma \gamma)$ and $b_\eta$.

\subsection{Three-point function}
\label{sec:supp-tseq-dependence}
The Euclidean amplitude in Eq.~\eqref{eq:eucl-amplitude} is accessed by evaluating the three-point function
\begin{equation} \label{eq:3pt-function}
\begin{aligned}
    C_{\mu\nu}(\tau,t_{\eta}) &\equiv \int d^3\vec{x} \, d^3\vec{y} \, e^{-i \vec{q}_1 \cdot \vec{x}} e^{i \vec{p} \cdot \vec{y}} \\
    &\quad \times \braket{T\{j_\mu(\tau,\vec x) j_\nu(0) \mathcal{O}_\eta^\dag(-t_{\eta}, \vec{y})\}}.
\end{aligned}
\end{equation}
For any operator $\mathcal{O}^\dag_\eta$ with overlap onto the $\eta$ state,
the three-point function is projected onto
the physical $\eta$ meson at large time separation, $-t_{\eta} \ll \min(0,\tau)$, irrespective of $\eta-\eta'$ mixing.
Here we use $\mathcal{O}_\eta^\dag = i \bar{\psi} \lambda_8 \gamma_5 \psi$,
where $\lambda_8 = \diag(1, 1, -2) / \sqrt{3}$ describes the $\operatorname{SU}(3)$ flavor structure. The validity of this choice and overlap onto the $\eta$ state are detailed in Appendix~\ref{app:interpolating-op-choice}. The electromagnetic currents are defined by $j_\mu = Z_V\,\bar\psi\,\gamma_\mu\,\mathcal{Q}\,\psi$
with $\mathcal{Q} = \mathrm{diag}(+2/3,\,-1/3,\,-1/3)$ and
$Z_V = 0.706378\,(16)$~\cite{ExtendedTwistedMass:2022jpw}.

Two remarks are in order concerning the definition
of the three-point function $C_{\mu\nu}$ using nonconserved
currents. First, one can show that 
potential short-distance singularities are absent in
Eq.~\eqref{eq:3pt-function} and that the
definition admits a well defined continuum limit. The argument is
given in Appendix D of Ref.~\cite{Gerardin:2016cqj} for Wilson
fermions and, by universality, applies to Wilson twisted-mass
lattice QCD as well. Second, we
note that the nonconserved currents do not spoil the
automatic ${\cal O}(a)$-improvement. This is because all involved
lattice quantities are constructed such that their parity
covariance is ensured, i.e., they have the correct symmetry property
under the 
twisted-mass parity transformation\footnote{Ordinary parity combined with a flavor exchange. See Ref.~\cite{Shindler:2007vp} for a comprehensive
listing of symmetries of the twisted-mass action.}. As a consequence, the
symmetry excludes the appearance of ${\cal O}(a)$
terms in physical matrix elements, as usual for twisted-mass
lattice QCD at maximal twist
\cite{Frezzotti:2003ni,Frezzotti:2004wz}, and hence guarantees automatic ${\cal
O}(a)$-improvement of the three-point function in
Eq.~\eqref{eq:3pt-function}.

\begin{figure}[t]
    \centering
    \includegraphics{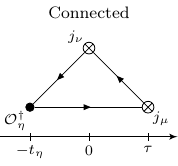}
    \includegraphics{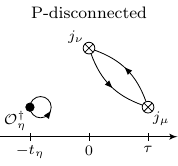}
    \includegraphics{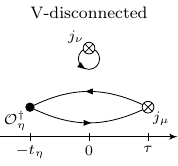}
    \includegraphics{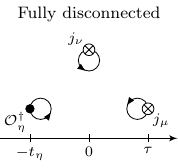}
    \caption{Wick contractions contributing to $C_{\mu\nu}(\tau,t_{\eta})$. The second connected diagram with quark propagators running in the opposite direction and the second V-disconnected diagram with a loop at $j_\mu(\tau)$ are omitted for brevity.
  \label{fig:3pt-wick-contractions}}    
\end{figure}

Evaluating $C_{\mu\nu}$ requires the Wick contractions shown in \fig~\ref{fig:3pt-wick-contractions}. 
We evaluate all connected (sub-)diagrams based on point-to-all quark propagators: we build the fully connected three-point function (top-left 
Wick diagram in \fig~\ref{fig:3pt-wick-contractions}) from a point-to-all propagator with spin-color diluted point sources
at the vertex labeled ``$j_\nu$'', with a subsequent sequential inversion through the $\mathcal{O}_{\eta}^{\dagger}$ vertex.
The sequential source for this inversion is the point-to-all propagator evaluated on timeslice $-t_{\eta}$, and multiplied 
by $\gamma_5$ to account for the pseudoscalar $\eta$-meson
interpolator. Since the $\eta$ meson is taken in its rest frame, no three-momentum 
is inserted in the sequential source.

In the P-disconnected diagram, we compute the quark-loop at $O_\eta^\dag$ from propagators based on stochastic volume sources.
Straightforward (undiluted) volume sources are sufficient in this case, and we ensure that the contribution from stochastic noise 
is suppressed below the noise from gauge configurations.

The connected current-current two-point function sub-diagram
(top-right Wick diagram in \fig~\ref{fig:3pt-wick-contractions})
we evaluate again using spin-color diluted point-to-all propagators, to allow for efficient computation with the large range
of photon three-momenta employed.

Unlike in previous lattice QCD studies of the $\pi^0$ TFF, here
P-disconnected diagrams of the isospin-singlet $\eta$-meson operator are nonzero.
The projection onto the $\eta$-meson state relies on a delicate cancellation between connected and P-disconnected diagram contributions, as shown in \fig~\ref{fig:3pt-decomposition}.

The amplitude $\ampl_{\mu\nu}$ is then recovered from $C_{\mu\nu}$ as
\begin{equation}
    \ampl_{\mu\nu}(\tau) = \lim_{t_{\eta} \rightarrow \infty} \frac{2 E_{\eta}}{Z_{\eta}} e^{E_{\eta} t_{\eta}} C_{\mu\nu}(\tau, t_{\eta}),
\end{equation}
where $Z_\eta = \braket{0 | \mathcal{O}_{\eta}(0,\vec{0}) | \eta(\vec{p})}$ is the overlap factor associated with the chosen creation operator.
In practice we approximate the limit $t_{\eta} \rightarrow \infty$ by considering three fixed values of $t_{\eta}$ in the range $0.80 \text{ fm} \lesssim t_{\eta} \lesssim 1.11 \mathrm{~fm}$.
Contamination from excited states and the $\eta'$ meson are suppressed best for the largest value of $t_\eta$, thus we report the values for $\Gamma(\eta\to\gamma\gamma)$, $b_\eta$, and $\amu{\eta}$
from $t_\eta \approx 1.11\,\mathrm{fm}$ as the main result and use the remaining choices 
to check for excited state effects.

Statistical noise significantly hinders evaluation of
$\ampl_{\mu\nu}(\tau)$ for large values of
$|\tau|$. Furthermore, the finite time extent of the lattice geometry would prevent integrating in the limits $\tau \rightarrow \pm \infty$ even if perfectly precise data were available. To address these issues,
following Refs.~\cite{Gerardin:2016cqj,Gerardin:2019vio,Burri:2021cxr}, we
perform joint fits of the asymptotic behavior of $\ampl_{\mu\nu}(\tau)$
for all $\vec{q}_1$ to Vector Meson Dominance and Lowest Meson Dominance functional forms~\cite{Chernyak:2014wra} with 
fit windows defined by $ t_i \le |\tau| \le t_f$. Details of the fitting procedure are described in Appendix.~\ref{app:tail-fits}
We then integrate over $\tau$ as in Eq.~(\ref{eq:tff-eucl-amplitude}) using numerical integration
of the lattice data within the peak region, $|\tau| \leq \tau_c$,
and analytical integration of the fit form in the tail region, $|\tau| > \tau_c$.
In this work, we consider several choices of $\tau_c$ in the range
$0.16\,\mathrm{fm} \lesssim \tau_c \lesssim
0.64\,\mathrm{fm}$. Variation between the results computed using
  different choices of $\tau_c$ gives a measure of the uncertainties
  resulting from noisy data in the tails and finite time extent effects.

\subsection{Extraction of \texorpdfstring{$Z_\eta$}{Zeta}  and \texorpdfstring{$E_\eta$}{Eeta}}

\begin{figure*}
    \centering
    \includegraphics{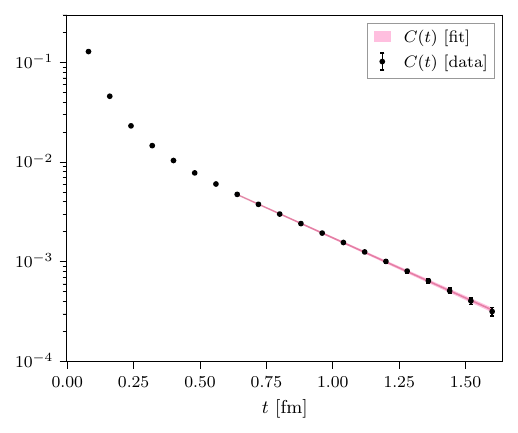}
    \includegraphics{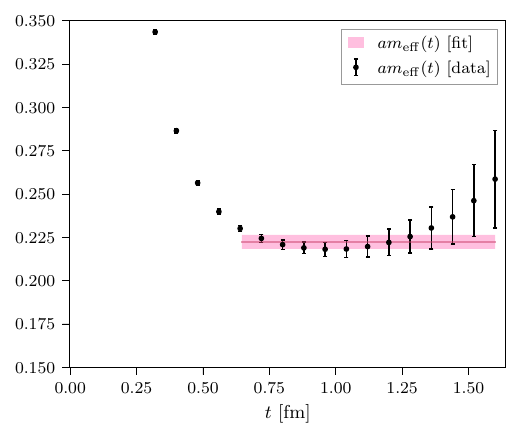}
    \caption{Two-point function $C(t)$ in lattice units and the corresponding effective mass $a m_{\mathrm{eff}}(t)$ versus fits used to extract the overlap and mass parameters.}
    \label{fig:2pt-meff}
\end{figure*}

The quantities $Z_\eta$ and $E_\eta = m_\eta$ (at rest) are extracted by fitting the two-point function of the interpolating operator selected above,
\begin{equation}
    C(t) \equiv a^3 \sum_{\vec{x}} \braket{ O_8(\vec{x},t) O^\dag_8(\vec{0},0) }.
\end{equation}
As the imaginary time separation $t$ is taken large, the asymptotic scaling of this function is given by a spectral decomposition,
\begin{equation}
\begin{aligned}
    C(t) \stackrel{t \rightarrow \infty}{\sim} \frac{a^4 |Z_\eta|^2}{2 a m_\eta} e^{-m_\eta t} + \text{excited states,}
\end{aligned}
\end{equation}
where the factor of $2 m_\eta$ is due to the relativistic normalization of the state $\ket{\eta}$ in the definition of $Z_\eta$.
As shown in Fig.~\ref{fig:2pt-meff}, we apply a two-state fit to accurately determine the scaling behavior of the two-point correlation function and its effective mass, $a m_{\mathrm{eff}}(t) \equiv -\log(C(t+a)) + \log(C(t))$, on the cB211.072.64 ensemble used in this work. The resulting overlap and mass parameters are determined in lattice units to be
\begin{align}
    a m_\eta &= 0.222(4), \\
    a^2 Z_\eta &= 0.112(3).
\end{align}
Using the lattice spacing $a=0.07957(13)$ fm determined in Ref.~\cite{ExtendedTwistedMass:2022jpw} this yields $m_\eta = 551.3(1.3)$ MeV in physical units. This is less than 8 permille higher than the experimental value and demonstrates the accuracy of our tuning of the valence strange-quark mass to reproduce the $\eta$-meson mass. Using alternative physical quantities, such as the $m_\Omega$ or $m_K$, yields differences between $6-11$\% supporting our expectation that lattice artifacts are subleading w.r.t.~the dominant statistical and other systematic errors in the TFF.

The two-point function is measured on the same gauge ensemble as the three-point function, and errors on these quantities are propagated through the calculation in a fully correlated way by using a per-bootstrap evaluation of the fitted quantities in each subsequent three-point analysis.

\subsection{Extrapolation via global conformal fit}
Access to the partial decay width, the slope parameter, and the $\eta$-pole HLbL contribution requires an interpolation
of the TFF close to the origin and an extrapolation in the quadrant of nonpositive photon virtualities.
We apply the model-independent expansion in powers of conformal variables advocated in Ref.~\cite{Gerardin:2019vio}, termed the $z$-expansion.
Analyticity of the form factor below the two-pion thresholds at $q_1^2 = 4 m_{\pi}^2$ and $q_2^2 = 4 m_{\pi}^2$ guarantees convergence as the highest power $N$ in the expansion is taken to infinity.
Moreover, the expansion is restricted to account for the known threshold scaling 
and contains preconditioning to more easily capture the expected
asymptotic behavior as $q_1^2, q_2^2 \rightarrow -\infty$.
In practice
we find that the $N = 2$ fit, consisting of six free parameters, already provides a very accurate fit to all lattice results, so we restrict to $N \in \{1, 2\}$ in all subsequent analyses.

To interpolate and extrapolate TFF data in the $(Q_1^2, Q_2^2)$ plane, we apply a global fit of the TFF data determined across all $\vec{q}_1$ using a model-independent $z$-expansion of order $N \in \{1, 2\}$. Variation between the choice of order is included in the model averaging of all final quantities as a systematic error.
The precise fit form used in this work is identical to the choice put forward in Ref.~\cite{Gerardin:2019vio}. For completeness, we review this approach here.

Noting that the TFF is analytic for all virtualities $Q_{1,2}^2 \geq -4m_\pi^2$ (including in particular the entire spacelike quadrant, $Q_{1,2}^2 \geq 0$), a conformal transformation is applied to yield the new variables
\begin{equation}
    z_k = \frac{\sqrt{t_c + Q_k^2} - \sqrt{t_c - t_0}}{\sqrt{t_c + Q_k^2} + \sqrt{t_c - t_0}}, \quad k \in \{1,2\},
\end{equation}
where $t_c = 4 m_\pi^2$ and $t_0$ is a free parameter that determines which virtualities are mapped to the origin of the new coordinates. In the resulting $(z_1, z_2)$ coordinates, the TFF is analytic for all $|z_{1,2}|^2 < 1$ and can be expanded in this domain as a polynomial in $z_{1,2}$, giving
\begin{equation} \label{eq:zexp-prefactor}
    \left( 1 + \frac{Q_1^2 + Q_2^2}{M_V^2} \right) \vvTFF{\eta}(-Q_1^2, -Q_2^2) = \sum_{n,m = 0}^{\infty} c_{nm} z_1^{n} z_2^{m},
\end{equation}
where Bose symmetry requires that $c_{nm} = c_{mn}$. In this expansion, the TFF is preconditioned to implement the known large-virtuality behavior already at zeroth order in the conformal expansion by including the prefactor $1 + (Q_1^2 + Q_2^2)/M_V^2$, where $M_V = 774\,\mathrm{MeV}$ is the vector-meson mass.

An order-$N$ truncation of the conformal expansion then provides a model-independent fit form to the TFF which must converge as $N \rightarrow \infty$. At finite $N$, it is useful to further restrict the coefficients $c_{nm}$ to enforce the appropriate scaling at threshold~\cite{Bourrely:2008za} by fixing the derivatives at $z_{1,2} = -1$ to zero, yielding the fit function
\begin{widetext}
\begin{equation} \label{eq:zexp-fit-function}
\begin{aligned}
    &\vvTFF{\eta}^{(z-\mathrm{exp},N)}(-Q_1^2, -Q_2^2) = \\
    & \quad \left( 1 + \frac{Q_1^2 + Q_2^2}{M_V^2} \right)^{-1} \sum_{n,m = 0}^{N} c_{nm} \left( z_1^{n}  - (-1)^{N + n + 1}\frac{n}{N+1} z_1^{N+1} \right) \left( z_2^{m}  - (-1)^{N + m + 1}\frac{m}{N+1} z_2^{N+1} \right)
\end{aligned}
\end{equation}
\end{widetext}
parameterized by $N(N+1)/2$ fit parameters $c_{nm} = c_{mn}$.

Finally, to optimize the rate of convergence to the TFF in the interval $-4 m_\pi^2 \leq Q_{1,2}^2 \leq Q_{\mathrm{max}}^2$, the parameter $t_0$ is chosen to be
\begin{equation}
    t_0 = t_c \left(1 - \sqrt{1 + Q_{\mathrm{max}}^2 / t_c} \right).
\end{equation}
In this work, we fix $Q_{\mathrm{max}}^2 = 4.0 \,\mathrm{GeV}^2$. Regardless of the choice of $Q_{\mathrm{max}}^2$, the $z$ expansion of the form given in Eq.~\eqref{eq:zexp-prefactor} is guaranteed to be valid by analyticity.

We then fit the parameters of the function in Eq.~\eqref{eq:zexp-fit-function} to our determined values of the TFF across all choices of ${\vec{q}_1}^{\,2}$ (the orbits shown in Fig.~2 in the main text) and for choices of $\omega_1$ selected per orbit to access virtualities $Q_{1,2}^2$ for which the ratios $Q_1^2/Q_2^2$ take values linearly interpolating between $0$ and $1$ along with the choices corresponding to exchanging $Q_1 \leftrightarrow Q_2$. In total, we evaluate $201$ choices of $\omega_1$ per orbit.

Data that correspond to identical momentum $\vec{q}_1$ and differ only in $\omega_1$ are strongly correlated, as the TFF for such choices differ only in the Laplace transform applied to identical lattice data. Data that correspond to distinct momenta $\vec{q}_1 \neq \vec{q}^{\;\prime}_1$ are also significantly correlated due to the common underlying gauge configurations and the global fit used in the integration of $\tilde{A}(\tau)$. This complicates estimation of the covariance matrix required for a
correlated fit. On the other hand, the model averaging procedure described in the following section is formulated to avoid needing estimates of the $\chi^2$ for fits. As such, throughout this work we choose to use uncorrelated $z$-expansion fits to the TFF data for all quantities.

The use of an uncorrelated fit means that the associated $\chi^2$ is an unreliable measure of goodness of fit. However, the quality of the fit at order $N=2$ can be seen in Fig.~5 of the main text, which shows that the conformal expansion already nearly interpolates the lattice data at all orbits using only $N(N+1)/2 = 6$ parameters. Thus only fits using orders $N \leq 2$ were considered in this work.

\subsection{Evaluation of \texorpdfstring{$\amu{\eta}$}{amu(eta-pole)}}
The $\eta$-pole HLbL contribution has the integral representation~\cite{Jegerlehner:2009ry,Nyffeler:2016gnb}
\begin{equation}
\begin{aligned}
  &\amu{\eta} = \left(\frac{\alpha}{\pi}\right)^{3} \int_0^{\infty} dQ_1 dQ_2 \int_{-1}^{1} dt \Big[  \\
  &\; w_1(Q_1, Q_2, t) \vvTFF{\eta}(-Q_1^2, -Q_3^2) \vrTFF{\eta}(-Q_2^2, \mathrlap{0)} \\
    &+ w_2(Q_1, Q_2, t) \vvTFF{\eta}(-Q_1^2, -Q_2^2) \vrTFF{\eta}(-Q_3^2, \mathrlap{0) \Big] \,,}
\label{eq:amu-eta-pole12}
\end{aligned}
\end{equation}
with $t=\cos \theta$ parameterizing the angle between the four-momenta, so that $Q_3^2 = Q_1^2 + 2 Q_1 Q_2 \cos \theta +  Q_2^2$.
The weight functions $w_1$ and $w_2$ are peaked such that contributions to Eq.~\eqref{eq:amu-eta-pole12} mainly come from the region $0 \leq Q_1, Q_2 \lesssim 2 \, \mathrm{GeV}$~\cite{Nyffeler:2016gnb}.
Knowledge of the TFF in the regime of relatively small virtualities
is thus sufficient to accurately evaluate $\amu{\eta}$.

Finally, we quantify systematic errors associated with the choices of tail-fit model, the parameters $(t_i, t_f)$, $\tau_c$ and the $z$-expansion order $N$ by the model-averaging procedure detailed in Appendix~\ref{app:error-estimation}. \\

\section{Results}
Our lattice results are obtained on the $2+1+1$ flavor gauge ensemble cB211.072.64 produced by the Extended Twisted Mass Collaboration (ETMC)~\cite{ExtendedTwistedMass:2021gbo}. Key features of this ensemble are given in Table~\ref{tab:cB-ens}.
The sea-quark masses for this ensemble are tuned to be isospin symmetric ($m_u = m_d$) and to reproduce the physical charged-pion mass and the strange- and charm-quark masses,
with a lattice spacing of $a \simeq 0.08 \mathrm{~fm}$ and a lattice size of $L \simeq 5.09 \mathrm{~fm}$ ($m_\pi L \simeq 3.62$)~\cite{ExtendedTwistedMass:2021gbo,ExtendedTwistedMass:2022jpw}.
The lattice spacing has been determined precisely in Ref.~\cite{ExtendedTwistedMass:2022jpw} using a combined analysis of meson observables across available ETMC ensembles to control finite-size effects and pion-mass dependence; in the present work, the uncertainty on the lattice-spacing determination is far below that of the lattice observables measured and these uncertainties are therefore neglected.
For the valence strange quark we use the mixed action approach in Ref.~\cite{Frezzotti:2004wz} with a valence strange-quark doublet, whose mass is 
tuned such that the $\eta$ meson has physical mass.

\begin{table*}
    \renewcommand{\arraystretch}{1.5}
    \begin{tabular}{l @{\hspace{0.5cm}} c @{\hspace{0.3cm}} c @{\hspace{0.3cm}} c @{\hspace{0.3cm}} c @{\hspace{0.3cm}} c}
    \toprule
         Ensemble & $L^3 \times T$ & MDUs & $a m_{\pi}$ & $m_{\pi} L$ & $m_{\pi}$ [MeV] \\
    \hline
         cB211.072.64 & $64^3 \times 128$ & $3161$ & $0.05659 (8)$ & $3.62$ & $136.8 (0.6)$ \\
    \botrule
    \end{tabular}
    \caption{Key details of the cB211.072.64 gauge ensemble used in this work.}
    \label{tab:cB-ens}
\end{table*}

All two-point and three-point function measurements were performed on a subset of $1539$ configurations separated by two MDUs each. 
For the evaluation of the connected Wick contractions of the three-point function,
we use $16$ point sources per configuration ($24624$ total inversions).
For the current-current two-point contraction in the $P$-disconnected diagram of the three-point function and for the connected two-point function measurements 
we use $200$ point sources per gauge configuration ($307800$ total inversions). Finally, we use $128$ stochastic sources per configuration ($196992$ total inversions) to evaluate pseudoscalar loops in the disconnected diagrams of both the three-point and two-point functions.
Due to the twisted-mass valence action for the light- and strange-quark doublet we can use the ``one-end-trick'' noise reduction technique for the
pseudoscalar, iso-scalar loops: In twisted-mass lattice QCD the iso-scalar loop (for either the light- or strange-quark doublet) is 
represented by chiral rotation as the difference of quark loops with positive and negative twisted quark mass. The latter difference
is converted into a two-point function with an additional sum over the lattice four-volume. This volume average leads to enhanced suppression of stochastic noise
and a more efficient stochastic estimator for the quark loop \cite{Alexandrou:2013wca}.

\begin{figure}
    \centering
    \vspace{-0.2cm}
    \includegraphics{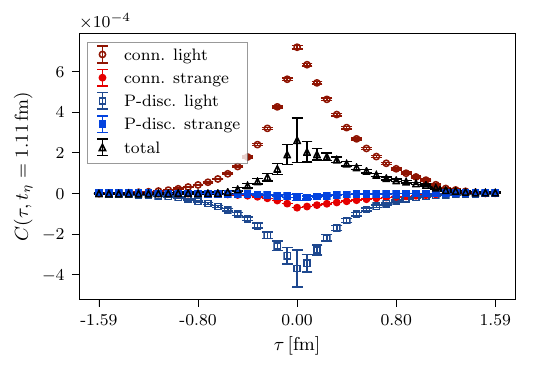}
    \vspace{-0.8cm}
    \caption{Contributions from the connected and P-disconnected Wick contractions 
    in the evaluation of the amplitude $C(\tau,t_{\eta}) \equiv
      (i a \eps_{ijk} \vec{q}_1^{\,i}/ |\vec{q}_1|^2) \,
      C_{jk}(\tau,t_{\eta})$ in lattice units
    at $t_\eta = 1.11\,\mathrm{fm}$ and $|\vec{q}_1^{\,2}| = 3 (2\pi/L)^2$.
    The labels ``light'' and ``strange'' indicate the quark flavor in the contractions of the electromagnetic currents.
    }
    \label{fig:3pt-decomposition}
  \end{figure}
We show in \fig~\ref{fig:3pt-decomposition} an example of the
contributions to $C_{\mu\nu}(\tau, t_{\eta})$ from the connected and
P-disconnected Wick contractions on this ensemble at our largest
separation, $t_{\eta} \simeq 1.11\,\mathrm{fm}$.
The contributions involving strange-quark vector currents are suppressed by a factor $\sim 10$ for the connected and $\sim 20$ for the P-disconnected contribution
compared to those from the light-quark vector currents. Contributions from charm-quark vector currents are expected to be even more suppressed,
as are those from V-disconnected and fully disconnected diagrams 
(lower two diagrams in \fig~\ref{fig:3pt-wick-contractions}), based on numerical evidence from recent results for the analogous pion TFF and for the $\eta$-meson TFF~\cite{Gerardin:2019vio,Burri:2021cxr,Gerardin:2023naa,Alexandrou:2023lia}.
At the presently achievable accuracy these contributions are hence not relevant and are not included in the analysis.

\begin{figure*}
    \centering
    \includegraphics{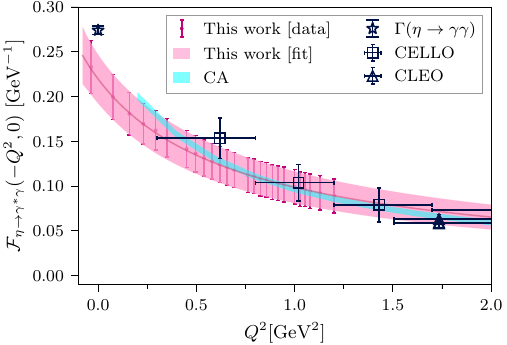}
    \includegraphics{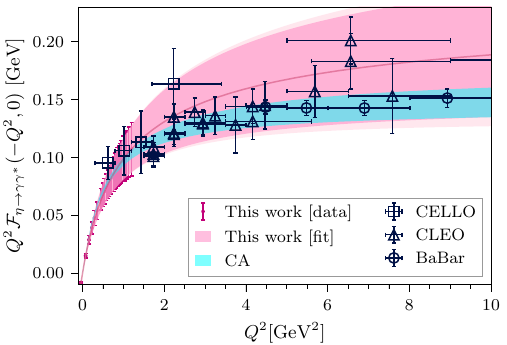}\\
    \vspace*{0.2cm}
    \includegraphics{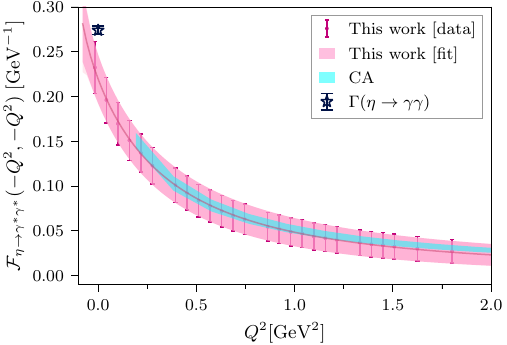}
    \includegraphics{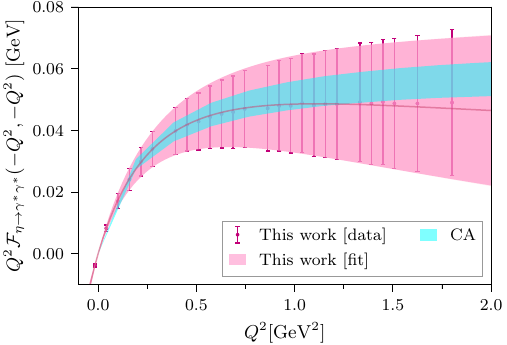}
    \caption{Comparison of the TFF estimated from this work
    (pink points corresponding to the accessible
    orbits shown in Fig.~\ref{fig:eta-orbits} and the pink curve showing the global conformal fit) versus the available $\vrTFF{\eta}$ and $\Gamma(\eta \to \gamma \gamma)$ experimental results (blue points)~\cite{CELLO:1990klc,CLEO:1997fho,BaBar:2009rrj,BaBar:2011nrp,Workman:2022ynf} and a Canterbury approximant estimate (cyan curve)~\cite{Masjuan:2017tvw}. Results from this work are based on a single lattice spacing and lattice volume, and the plotted uncertainties thus exclude lattice discretization and finite-size effects which will be studied in future work. For better comparison to features at both small and large $Q^2$, the TFFs are plotted both with and without a conventional $Q^2$ prefactor.
    \label{fig:TFF-comparison}
    }
\end{figure*}

In \fig~\ref{fig:TFF-comparison} we show our results for the TFF as a
function of the virtuality in the single-virtual case
$\vrTFF{\eta}(-Q^2,0)$ (top row) and in the double-virtual case
$\vvTFF{\eta}(-Q^2,-Q^2)$ (bottom row) together with our result from
the $z$-expansion fits.
The darker inner band indicates only statistical uncertainties while the lighter outer band includes systematic uncertainties estimated from the variation of fitting choices discussed above. At all virtualities shown, the statistical errors dominate the total uncertainty.
In addition to the available experimental data, we also show the Canterbury approximant (CA) result from Ref.~\cite{Masjuan:2017tvw}.
We observe reasonable agreement between our results, the experimental data and the CA data.

From the parameterization of the momentum dependence of our TFF data we extract the decay width, slope parameter, and $\amu{\eta}$.
As with the TFF itself, we repeat the calculation for
all choices of the analysis parameters
to determine systematic errors associated with tail fits of $\ampl$ and the $z$-expansion. A detailed breakdown is given in App.~\ref{app:error-estimation}.
For the decay width the resulting systematic uncertainty stems mainly from the variation in the fits of the tails of $\ampl_{\mu\nu}(\tau)$ and $\tau_c$, while for the slope parameter and the HLbL pole contribution it is mainly due to the conformal fit. The total error, however, is always dominated by the statistical uncertainties. We also observe a mild systematic dependence on $t_{\eta}$, as detailed below, which points to the fact that excited-state and possibly $\eta'$-meson contributions to the transition amplitude are not completely eliminated at the smaller values of $t_{\eta}$. We conservatively quote results obtained at our largest value of $t_{\eta} \simeq 1.11\,\mathrm{fm}$ for which the statistical uncertainty is largest and covers the results at the smaller $t_{\eta}$ values.

\begin{figure*}
    \centering
    \includegraphics{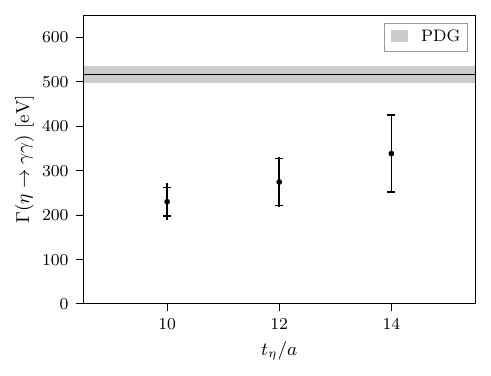}
    \includegraphics{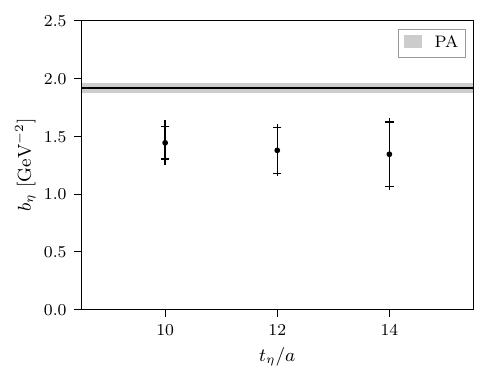}\\
    \includegraphics{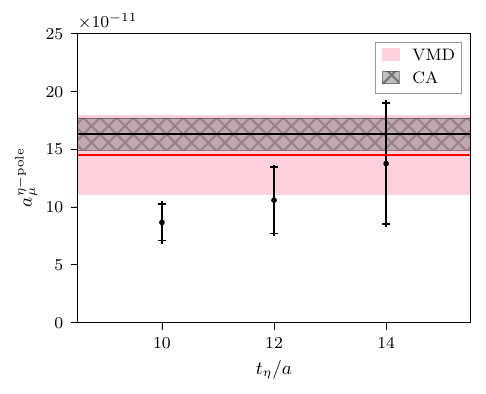}
    \caption{Comparison of the partial decay width
      $\Gamma(\eta\rightarrow\gamma\gamma)$, the slope parameter
      $b_\eta$, and  the $\eta$-pole contribution $\amu{\eta}$ from
      three choices of $t_{\eta}/a=10, 12, 14$ corresponding
      to $t_\eta=0.80, 0.96, 1.11 \,\mathrm{fm}$. For reference, the values are respectively compared against estimates from the PDG~\cite{Workman:2022ynf}, Pad\'{e} approximant (PA) fits to experimental data~\cite{Escribano:2015nra}, and the VMD model~\cite{Nyffeler:2016gnb} and Canterbury approximant (CA) experimental fits~\cite{Masjuan:2017tvw}.
      Results from this work are based on a single lattice spacing and lattice volume, and the plotted uncertainties thus exclude lattice discretization and finite-size effects which will be studied in future work.}
    \label{fig:combined-tseq-comparison}
\end{figure*}

In Fig.~\ref{fig:combined-tseq-comparison} we
show the dependence of the partial decay width
$\Gamma(\eta\rightarrow\gamma\gamma)$, the slope parameter $b_\eta$,
and the $\eta$-pole contribution $\amu{\eta}$ on the
choice of $t_\eta$ which denotes the imaginary time location of the creation operator $\mathcal{O}^\dag_\eta(-t_\eta)$ for the $\eta$ meson, to be compared with imaginary time coordinates of the currents $j_\mu(\tau)$ and $j_\nu(0)$.  The outer error bar denotes the total error,
while the inner one shows the statistical error only. It is clear that
the total error is dominated by the statistical one in all
cases and for all $t_\eta$ considered in this calculation. For all three quantities we observe a
mild systematic trend with growing $t_\eta$ which may be an indication that excited state and $\eta'$-meson
contributions to the transition amplitude, and hence to the quantities
shown here, may still be present at the smaller values of
$t_\eta$. Since we are interested in the limit $t_\eta \rightarrow
\infty$ we conservatively quote the results for the largest available
$t_\eta$ for which the statistical error is largest and covers the
results at the smaller values of $t_\eta$.

For the leading-order decay width we obtain 
\begin{equation} \label{eq:Gamma-result}
    \Gamma(\eta \to \gamma \gamma) =\Gammaresult \,\mathrm{eV}
\end{equation}
in comparison to the experimental average $516(18)\,\mathrm{eV}$~\cite{Workman:2022ynf,JADE:1985biu,CrystalBall:1988xvy,Roe:1989qy,Baru:1990pc,KLOE-2:2012lws}.
For the slope parameter we find
\begin{equation} \label{eq:b-eta-result}
    b_\eta = \bPresult \, \mathrm{GeV}^{-2}
\end{equation}
to be compared with $b_\eta=\bPresultExpt$ GeV$^{-2}$ from a Pad\'{e} approximant fit to the experimental results \cite{Escribano:2015nra} and $b_\eta=\bPresultDisp$ GeV$^{-2}$ from a dispersive calculation~\cite{Kubis:2015sga}.
Finally, we use the parameterization of our TFF data to perform the integration in Eq.~\eqref{eq:amu-eta-pole12} and obtain
\begin{equation}
\amu{\eta} = \amuresultshort \cdot 10^{-11} 
\label{eq:amueta-res}
\end{equation}
in comparison to a Canterbury approximant fit to experimental results yielding $16.3(1.4) \cdot 10^{-11}$~\cite{Masjuan:2017tvw}, the VMD model value $14.5(3.4) \cdot 10^{-11}$~\cite{Nyffeler:2016gnb}, and estimates $15.8(1.2) \cdot 10^{-11}$~\cite{Eichmann:2019tjk} and $14.7(1.9) \cdot 10^{-11}$~\cite{Raya:2019dnh} based on the Dyson-Schwinger equations.

We emphasize that our results are obtained at a fixed lattice spacing and a fixed volume.
The present estimates therefore exclude systematic errors associated
with finite-volume effects and lattice artifacts. 
The latter are expected to be of $\mathcal{O}\left(a^2 \Lambda_\text{QCD}^2\right)$
with the lattice discretization used here, while the former are
expected to be suppressed by $\exp(-m_\pi L)$ with $m_\pi L \simeq3.62$.
They are hence expected to be subleading w.r.t.~to the dominating statistical and
other systematic errors in the TFF. Lattice artifacts contribute through the bare TFFs, the vector-current renormalization factors (except in $b_\eta$) 
and through the setting of the lattice scale required to convert $m_{\mu}$ to lattice units. 
Both $Z_V$ and the lattice scale are determined 
independently of the quantities considered here~\cite{ExtendedTwistedMass:2022jpw,ExtendedTwistedMass:2021gbo}.
A quantitative estimate of the lattice artifacts present in
$\amu{\eta}$ can therefore be obtained by considering the
scheme of fixing the renormalization by the physical decay width instead of the hadronic scheme.
This gives
$a_{\mu;\Gamma-\mathrm{renorm}}^{\eta\text{-pole}} = 20.7 \, (4.5)_{\mathrm{stat}} (2.3)_{\mathrm{syst}} \cdot 10^{-11}$,
which differs from $\amu{\eta}$ in Eq.~(\ref{eq:amueta-res}) by $6.9
\cdot 10^{-11}$ and is of similar size as our total error. \\

\section{Conclusions and outlook}
The results of our lattice QCD calculation of the transition form factor $\vvTFF{\eta}(q_1^2,q_2^2)$ at physical pion mass 
have a precision comparable to experimental results in the range
where both are available, and demonstrate nice agreement, cf.~\fig~\ref{fig:TFF-comparison}.
Our results provide single-virtual data at lower photon virtuality than currently accessible by experiments.
This includes the region around zero virtuality necessary to study the decay width and slope parameter.
The results for these
quantities in Eqs.~\eqref{eq:Gamma-result} and \eqref{eq:b-eta-result} undershoot the experimental (and for $b_\eta$ also theoretical)
results by 1.5--2.0 standard deviations.

Our lattice computation also provides TFF data for double-virtual (space-like) photon kinematics, which is difficult to access by experiment.
We have made use of this advantage and calculated the $\eta$-pole contribution to the anomalous magnetic moment of the muon, $\amu{\eta} = \amuresultshort \cdot 10^{-11}$.
Our result confirms the currently available data-driven Canterbury approximant estimate~\cite{Masjuan:2017tvw} and the theoretical model estimates~\cite{Nyffeler:2016gnb,Eichmann:2019tjk,Raya:2019dnh},
but does not yet reach the same precision.
Nevertheless, it provides important independent support of these estimates.
The main shortcoming of our calculation is the use of a single lattice spacing, which
will be removed in the future by computations with ETMC gauge ensembles on finer lattices~\cite{ExtendedTwistedMass:2021gbo,ExtendedTwistedMass:2021qui}.

Note added: While our paper was under review a comprehensive study
  of the pseudoscalar TFFs and their contribution to $a_\mu$  has
  appeared, including results for the $\eta$ meson \cite{Gerardin:2023naa}.

\begin{acknowledgments}
We thank Martin Hoferichter, Simon Holz, and Bastian Kubis for helpful discussions. 
We are also grateful to the authors of Ref.~\cite{Masjuan:2017tvw} for sharing form factor data produced in their work.
This work is supported in part by the Sino-German collaborative research center CRC 110 and
the Swiss National Science Foundation (SNSF) through grant No.~200021\_175761, 200020\_208222, and 200020\_200424.
We gratefully acknowledge computing time granted on Piz Daint at Centro Svizzero di Calcolo Scientifico (CSCS)
via the projects s849, s982, s1045 and s133.
Some figures were produced using \texttt{matplotlib}~\cite{Hunter:2007}.
\end{acknowledgments}

\appendix

\section{Error estimation and model averaging}
\label{app:error-estimation}
All statistical errors reported in this work are given as $1\sigma$ confidence intervals derived from $N_{\mathrm{boot}} = 2000$ bootstrap resamplings of the ensemble of configurations. We find virtually no autocorrelation between the relevant primary data taken on a subset of configurations constituting the ensemble, and the bootstrap bin size is therefore fixed to $1$.

\begin{table*}
    \centering
    \begin{tabular}{@{\hspace{0.5cm}}l c c c}
        \toprule
        & \hspace{0.2cm} $10^{11} \cdot \amu{\eta}$ \hspace{0.2cm}
        & \hspace{0.2cm} $\Gamma(\eta \to \gamma \gamma)$ [$\mathrm{eV}$] \hspace{0.2cm}
        & \hspace{0.2cm} $b_\eta$ [$\mathrm{GeV}^{-2}$] \hspace{0.2cm} \\
        \hline
        Tail model vs data cut ($\tau_c$) & 0.22 & 10.1 & 0.020  \\
        Tail fit windows ($t_i$, $t_f$) & 0.18 & 6.5 & 0.009  \\
        Fit model (VMD vs.\ LMD) & 0.31 & 11.6 & 0.034  \\
        Conformal fit order ($N$) & 1.44 & 1.8 & 0.123  \\
        \hline
        Total systematic & 1.53 & 17.2 & 0.135  \\
        \hline
        Statistical & 5.24 & 86.7 & 0.279  \\
        \hline
        \hline
        Total & 5.46 & 88.4 & 0.310  \\
        \botrule
    \end{tabular}
    \caption{Decomposition of uncertainties in the reported values of the three quantities studied at the single lattice spacing and volume used in this work. The results and uncertainties are based on the conservative choice $t_\eta/a = 14$ corresponding to $t_\eta = 1.11$ fm.}
    \label{tab:uncertainty-decomp}
\end{table*}

During our analysis, we make several choices corresponding to fits of the large-$|\tau|$ tails of the amplitude $\ampl_{\mu\nu}(\tau)$ and of the finite-volume TFF orbits. In particular, the following analysis parameters are varied:
\begin{enumerate}
    \item The choice between using the Vector Meson Dominance (VMD) or Lowest Meson Dominance (LMD) model to the fit the tail behavior;
    \item The window $(t_i, t_f)$, determining which regions of the amplitude $\ampl_{\mu\nu}(\tau)$ are used as inputs to fit the asymptotic tail behavior;
    \item The integration cutoff $\tau_c$, distinguishing the region $|\tau| \leq \tau_c$ in which the lattice data is integrated from the region $|\tau| > \tau_c$ in which the analytical tail model is integrated; and
    \item The order $N$ of the conformal expansion used to fit the TFFs.
    \end{enumerate}
The variation of our estimates with these model choices gives estimates of the systematic errors associated with these steps. We apply the approach of Refs.~\cite{doi:10.1126/science.1257050,Borsanyi:2020mff} to construct cumulative distribution functions (CDFs) of all final quantities with various subsets of models and with two choices of rescaling parameter $\lambda$ applied to the systematic error. The various total error estimates, given by the difference between the 16th and 84th percentiles of the CDF in each case, allow an extraction and decomposition of the total uncertainty into statistical, total systematic, and various individual sources.

In this approach, weights must be assigned to each model included in the CDF. Weights based on the Akaike Information Criterion~\cite{Akaike1978} derived from $\chi^2$ values of each fit have been employed in previous work. For the tail of the amplitude, we perform a fit to values of $\ampl_{\mu\nu}(\tau)$ over sequential choices of $\tau$ and across all momentum orbits. For the $z$-expansion, we perform a fit to values of $\vvTFF{\eta}(-Q_1^2, -Q_2^2)$ across all orbits at several fixed choices of the ratio $Q_1^2/Q_2^2$. As discussed in the previous section, this input data is highly correlated, and determining the correlated $\chi^2$ therefore requires a very precise estimate of nearly degenerate covariance matrices of both the tail fits and $z$-expansion fits. Even for fits to small windows $(t_i, t_f)$ and few choices of orbits, we found estimates of the $\chi^2$ values to be inaccurate and unstable in our preliminary investigations. Instead, in this work we derive all results from much more stable uncorrelated fits. For the model averaging, we then make the conservative choice to use a uniform weighting of all possible models in the CDF method. This can be expected to overestimate the systematic error associated with model variation.

The decomposition of uncertainties is detailed in Table~\ref{tab:uncertainty-decomp} for all three final physical quantities studied in this work. Due to correlations between the total error estimates in each case, the decomposition does not simply add in quadrature, but nevertheless gives an estimate of which components of the error dominate the error budget. Unsurprisingly, the dominant sources of systematic errors vary depending on the observable considered. For the $\eta$-pole contribution to the HLbL, the biggest source of systematic error is the conformal fit used to extrapolate the TFF $\vvTFF{\eta}(q_1^2, q_2^2)$ from the low-virtuality orbits accessible on the lattice to the full plane of spacelike $(q_1^2, q_2^2)$. This indicates that, despite the important contributions to $\amu{\eta}$ from low virtualities, the large uncertainties in the nearly unconstrained higher virtualities can still affect the estimate of $\amu{\eta}$ from lattice data alone. Incorporating some information about asymptotic scaling of the TFF at large virtualities is therefore an interesting prospect for future work. The other two quantities, $\Gamma(\eta \to \gamma \gamma)$ and $b_\eta$ are directly related to the behavior of the TFF at $q_1^2 = q_2^2 = 0$. In the case of $\Gamma(\eta \to \gamma \gamma)$, the choices used to fit the tails of the amplitude $\ampl_{\mu\nu}(\tau)$ dominate the systematic errors, while for $b_\eta$ the systematic uncertainties are still set by the conformal expansion fit. Nonetheless, we find that the uncertainties in all three quantities are almost entirely given by the statistical error, which always far outweighs the systematic errors.

\begin{figure}
    \centering
    \includegraphics[width=\linewidth]{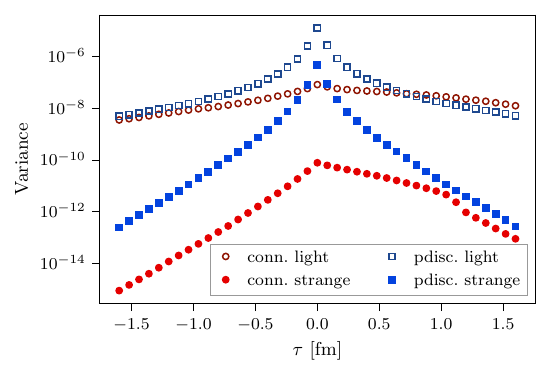}
    \caption{Comparison of the variance independently evaluated for each Wick contraction contributing to $C(\tau,t_{\eta})$ at $t_\eta = 1.11\,\mathrm{fm}$ and $|\vec{q}_1^{\,2}| = 3 (2\pi/L)^2$.}
    \label{fig:3pt-err-decomposition}
\end{figure}

The global fit used in the integration of $\tilde{A}(\tau)$ prevents decomposing the precise contribution of statistical errors to the final values of $\amu{\eta}$, $\Gamma(\eta \to \gamma \gamma)$, and $b_\eta$. However, one can consider the relative contributions of various Wick contractions to $\tilde{A}(\tau)$ itself to qualitatively understand the dominant source of statistical error. This is shown for the example of the orbit $|\vec{q}_1^{\,2}| = 3 (2\pi / L)^2$ in Fig.~\ref{fig:3pt-err-decomposition}, which can be compared against the plot of these same contributions in Fig.~4 of the main text. Correlations of the errors prevent interpreting the contributions as a direct decomposition of the total error, however one can still identify the Wick contractions dominating the error for various values of $\tau$. In particular, at values of $|\tau| \lesssim 0.5\,\mathrm{fm}$, the P-disconnected diagrams dominate the variance, while for $|\tau| \gtrsim 0.5\,\mathrm{fm}$ the connected light diagram also makes a notable contribution.

\section{Interpolation of the \texorpdfstring{$\eta$}{eta} state}
\label{app:interpolating-op-choice}
The $\eta$-meson state is the lowest-lying eigenstate of the twisted-mass lattice Hamiltonian in the channel with quantum numbers $I^G\,\left(J^{PC}\right) = 0^{+}\,\left(0^{-+}\right)$. 
The exact interpolating field to project onto the $\eta$ eigenstate in the lattice calculation is unknown. However, it is sufficient that it can be written
as a linear combination of the quark-model octet- and singlet-pseudoscalar operators
\begin{equation}
\begin{aligned}
  O_{\eta}^{\mathrm{exact}} &= \alpha \, \bar{\psi} \lambda_8 \gamma_5 \psi + \beta \, \bar{\psi} \gamma_5 \psi + \dots \\
  &= \alpha \, \frac{1}{\sqrt{3}} \,\left( \bar u\gamma_5 u + \bar d\gamma_5 d - 2 \bar s \gamma_5 s  \right) \\
  &\quad + \beta \,\left( \bar u\gamma_5 u + \bar d\gamma_5 d + \bar s\gamma_5 s  \right) + \dots \,,
  \label{eq:eta-1}
\end{aligned}
\end{equation}
where the ellipsis denotes further linearly independent operators. Using the octet operator 
\begin{equation}
\begin{aligned}
  O_8 &= i \bar{\psi} \lambda_8 \gamma_5\psi = \frac{i}{\sqrt{3}} \,\left( \bar u\gamma_5 u + \bar d\gamma_5 d - 2 \bar s \gamma_5 s  \right)
  \label{eq:eta-2}
\end{aligned}
\end{equation}
as the interpolating operator means that the projection is imperfect, i.e., the creation operator will produce a tower of Hamiltonian eigenstates from the vacuum,
\begin{align}
  O_8^\dagger\,|\,0\rangle &= Z_{\eta}\,|\eta\rangle
  + Z_{\eta^\prime}\,|\eta^\prime\rangle + \dots \,,
  \label{eq:eta-3}
\end{align}
with increasing mass or energy and with $Z_{\eta} = \braket{0\,|\,O_8(0)\,|\,\eta}$, $Z_{\eta^\prime} = \braket{0\,|\,O_8(0)\,|\,\eta^\prime}$.
Nevertheless, the $\eta$-meson state is the unique ground state of lowest mass, and propagation in Euclidean time systematically suppresses the contribution of the $\eta^\prime$-meson and excited states 
lying higher in the spectrum. This suppression scales exponentially as $\exp\left( -(M - m_{\eta}) t \right)$, in terms of the Euclidean time evolution $t$ and the relative energy gap between the mass $M$ of the higher state and $m_\eta$. This applies to
all two- and three-point correlation functions used in this work. Thus for sufficiently long Euclidean time propagation, the projection onto the $\eta$-meson state is achieved 
by our choice of $O_8^\dag$ as the creation operator for the two-point and three-point functions.

\section{VMD and LMD fits to the amplitude}
\label{app:tail-fits}
As discussed in Sec.~\ref{sec:supp-tseq-dependence}, we perform global fits to the amplitudes $\tilde{A}_{\mu\nu}(\tau)$ across all vector current momenta $\vec{q}_1$ and use the resulting functional forms instead of data when integrating Eq.~\eqref{eq:tff-eucl-amplitude} at large $|\tau|$. Here we detail the functional forms used for the fits, which are inspired by the Vector Meson Dominance (VMD) and Lowest Meson Dominance (LMD) models~\cite{Moussallam:1994xp,Knecht:1999gb}.

The transition form factor in the VMD and LMD models are respectively given by
\begin{equation}
    \vvTFF{\eta}^{\mathrm{VMD}}(q_1^2, q_2^2) = \frac{\alpha M_V^4}{(M_V^2 - q_1^2)(M_V^2 - q_2^2)}
\end{equation}
and
\begin{equation}
    \vvTFF{\eta}^{\mathrm{LMD}}(q_1^2, q_2^2) = \frac{\alpha M_V^4 + \beta (q_1^2 + q_2^2)}{(M_V^2 - q_1^2)(M_V^2 - q_2^2)},
\end{equation}
where phenomenology suggests the particular choice $M_V = 775\,\mathrm{MeV}$ (the mass of the $\rho$ meson) and choices of $\alpha$ and $\beta$ to respectively match the triangle anomaly, which determines $\rrTFF{\eta}(0, 0)$ to leading order~\cite{Adler:1969gk,Bell:1969ts}, and the short distance doubly virtual behavior~\cite{Lepage:1979zb,Lepage:1980fj,Nesterenko:1982dn,Novikov:1983jt}. Note that the VMD model is simply a special case of the LMD model with $\beta$ fixed to zero. For fits to the lattice amplitude data, these parameters will be taken as free parameters of the fitting function.

Inverting the relation in Eq.~\eqref{eq:tff-eucl-amplitude} between the TFF and amplitude $\tilde{A}_{ij}(\tau)$ in the rest frame of the $\eta$ meson results in a functional form for the amplitude using the LMD model (or by fixing $\beta = 0$ the VMD model),
\begin{equation}
\begin{aligned}
\label{eq:A_tilde_fitfunc}
    \tilde{A}^{\mathrm{LMD}}_{ij}(\tau) &= 
    -i m_\eta \epsilon_{ijk} q_1^k e^{m_\eta |\tau| \Theta(-\tau)} \\
    &\qquad \times [ C_+ e^{-E_V |\tau|}
    - C_- e^{-(m_\eta + E_V)|\tau|}],
\end{aligned}
\end{equation}
where
\begin{equation}
  \begin{aligned}
    C_{\pm} &\equiv \frac{\alpha M_V^4 + \beta (2 M_V^2 + m_\eta^2 \mp 2 m_\eta E_V)}{m_\eta E_V(2E_V \mp m_\eta)}, \\
    E_V &\equiv \sqrt{M_V^2+|\vec{q}_1|^2}.
  \end{aligned}
\end{equation}
\vspace{1cm} 

\bibliographystyle{h-physrev5}
\bibliography{main}

\end{document}